\documentclass{emulateapj}
\usepackage{graphicx,color}
\usepackage{amssymb,amsmath}
\definecolor{mycolor1}{rgb}{0.1, 0.188, 0.478}
\usepackage[colorlinks,hyperfootnotes=false,citecolor=mycolor1,linkcolor=mycolor1]{hyperref}

\begin{document}

\shorttitle{Real-Time Cosmology with Repeating FRBs}
\shortauthors{Zitrin \& Eichler}

\slugcomment{Submitted to the Astrophysical Journal}

\title{Observing Cosmological Processes in Real Time with Repeating Fast Radio Bursts}
\author{Adi Zitrin\altaffilmark{*,$\dagger$} \& David Eichler\altaffilmark{*}}
\altaffiltext{*}{Dept. of Physics, Ben-Gurion University, Be'er-Sheva 84105, Israel}
\altaffiltext{$\dagger$}{adizitrin@gmail.com}

\begin{abstract}
It is noted that the duration of  a fast radio burst (FRB), about $10^{-3}$ s, is a smaller fraction of the time delay between multiple images of a source gravitationally lensed by a galaxy or galaxy cluster than the human lifetime is to the age of the universe.  Thus repeating, strongly lensed FRBs may offer an unprecedented opportunity for observing cosmological evolution in ``real time''. The possibility is discussed of observing cosmic expansion, transverse proper motion, mass accretion and perhaps growth of density perturbations, as a function of redshift.\\
\end{abstract}

\section{Introduction}\label{intro}

Fast Radio Bursts (FRBs) occur at an all sky rate that has been estimated to be $2 \cdot 10^3$ per day 
[$7.3  \cdot 10^5$ per year] above a flux density threshold of 1 Jy \citep[or fluence of 2 Jy ms;][]{KeanePetroff2015}, and $\gtrsim 10^{7.5}$ per year at 0.8 mJy (e.g. \citealt{FialkovLoeb2017}), representing the expected sensitivity of the Square Kilometer Array.  Because of limited sky coverage to date, only about 31 FRBs have been reported at the time of this writing, almost all of them detected over a 5 year period. On the other hand, if the all sky rate were $7 \cdot 10^5$ per year above 1 Jy ($2 \cdot 10^3$ per day, \citealt{KeanePetroff2015}), the sky-time coverage (i.e., the fraction of sky observed, times observing time) can be assumed to have been of the order of $4.3 \cdot 10^{-5}$~yr. If we assume for the sake of simplicity that all FRBs to date were found over an effective observing time of 5 years (1 year), then the fraction of covered sky is $4.3/5 \cdot 10^{-5} \simeq0.9 \cdot 10^{-5}$ ($4.3 \cdot 10^{-5}$).
Now of these 31, one FRB (FRB 121102) source repeated. Its redshift is 0.19 and it has a typical peak flux of 0.9 Jy. Given the above estimated sky coverage, it may be assumed, unless FRB 121102 is statistically very improbable  {\it a priori}, that $N_{1Jy} \sim 10^{4}-10^{5}$ such repeating FRBs could be eventually found over the entire sky above 1 Jy.

In practice, typical field-of-view sizes might only allow for discovering a small fraction of the repeating FRBs per year. On the other hand, it should be much easier to scan the sky for repeating FRBs than for non-repeating ones, depending on the frequency of repetition, which is still largely unknown.
   
Let us note that the equation of motion for an approaching photon is given by
 \begin {equation}
 Rdr = -cdt
 \end{equation}
where $c$ is the speed of light, $t$ is cosmic time, $r$ is the dimensionless comoving distance of the photon, and $R(t)$ is the cosmic scale factor so that $Rdr$ is the proper distance.  But $dt$ also obeys 
 \begin{equation}
 H(z) dt = dR(z)/R(z)  =- d(1+z)/(1+z),
 \end{equation}
 where $z$ is redshift and $H(z)$ is the Hubble constant at the time of emission.  It follows from the above that sources seen at present obey 
 \begin{flalign}\label{drdz}
  &d r(z)/d(1+z) = [dr/dt]/[d(1+z)/dt] \nonumber \\
  &= c/[(1+z)R(z)H(z)]  = c/[R_o H(z)]  \nonumber\\
  &= c/[R_o [(aR^{-3}+\Lambda)c^2/3]^{\frac{1}{2}}] \nonumber \\
  &= c/[R_o [(aR_o^{-3}(1+z)^3+\Lambda)c^2/3]^{\frac{1}{2}}]
 \end{flalign} 
   where $aR_o^{-3}=[\Omega_{m,0}/\Omega_{\Lambda,0}]\Lambda \simeq [0.3/0.7]\Lambda$, $\Lambda$ is the cosmological constant, $\Omega_{m,0}$ is the present cosmic matter density relative to the critical density and $\Omega_{\Lambda,0}$ is the present dark energy density relative to the critical density. Integrating equation (\ref{drdz}) yields $r(z)$, and, in a flat $\Lambda$CDM  model, the comoving volume of a sphere out to $r(z)$ is $4\pi r(z)^3/3$. 
   The total number of repeating FRBs expected is $N_{\epsilon}= \int_0^{\infty} (dn/dz)dz= (dn/dr^3)\int_0^{\infty} (dr^3/dz)dz $, where $dn/dr^3$, the volume density of the sources, is assumed to be constant. The total number of repeating sources that can be detected down to $\sim$0.8 mJy is given by cutting off the integral at the redshift at which a source at $z=0.19$ with peak flux $\sim$1 Jy  will appear $(1/0.8)\times 10^3$ times dimmer. This occurs at $z\simeq3.7$. Thus:
   \begin{equation}
   N_{\epsilon}\sim{1\over 2}\left[\frac{r(z=3.7)}{r(z=0.19)}\right]^3 N_{1 Jy} \sim 360 N_{1 Jy} \sim 10^{6.5} -10^{7.5} ,
   \end{equation}
where we have assumed that the comoving number density of repeating FRBs is about half of the inverse comoving volume subtended by a sphere at the redshift of FRB 121102, we have counted only those FRBs that are at least as intrinsically bright as FRB 121102, and we have assumed all such sources could be seen out to high redshift. This is roughly compatible with the estimate of \citet{FialkovLoeb2017} for all FRBs, and it would not be surprising if a sizable fraction of all FRBs repeat even many times per year, and that the closest are the ones whose repeating nature is the first to be observed. If the FRB rate scales in proportion to the star formation rate, then it was even higher at large redshift in analogy to the gamma ray burst rate \citep{EichlerGuettaPohl2010}.  Note that we are counting sources, not FRBs. So the repetitive nature would imply a still higher FRB rate. 
 
Now the strong-lensing cross section for a given typical rich cluster can comfortably be as high as 10 arcsec$^2$, or about $10^{-6}$ square degrees. But there are $\sim 10^4$ rich clusters in the sky with that cross section on average (e.g. \citealt{Zitrin2012UniversalRE}, \citealt{OguriBlandford2009}) so the chances are that a few repeating FRBs will be lensed by {\it some} rich cluster, and there should be no confusion between independent FRBs and repeating ones. The expected rate of repeating FRBs lensed by {\it individual galaxies} is far higher, and could be as high as  $10^4$ per sky per year.
 
Observing cosmic processes in real time is a long pursued idea in  cosmology. Our knowledge of the underlying cosmology has been revolutionized in recent decades thanks to, e.g., measurements of Type Ia supernovae \citep{Riess1998,Perlmutter1999}, baryonic acoustic oscillations \citep{Eisenstein2005}, time-delay measurements of lensed quasars \citep{Suyu2013_2,Tewes2013}, and the cosmic microwave background \citep{Planck2015Params}. However, these measurements generally -- or mostly -- constitute indirect probes of the cosmological parameters. Real-time cosmology refers to \emph{directly} measuring the change in radial and transverse location of cosmological sources over relatively
short periods of time, in different redshifts (\citealt{Quercellini2012} for a review), charting the actual growth or expansion history of the universe. While its principle has been proposed already several decades ago \citep{Sandage1962}, measurements are still unrealized given the long timescale needed to observe most phenomena compared to the human lifetime (see also \citealt{Lake1981}, and references therein). \citet{Loeb1998} suggested a couple of interesting ways to potentially measure the effect; one is via two observations,  set a decade apart, of large samples of quasars using sensitive high-resolution spectrographs where cross-correlating the Lyman-alpha absorption lines could reveal by how much the universe expanded in that time period (see also \citealt{Liske2008,Pasquini2010codex,Kloeckner2015arxiv,Martins2016}). A second idea suggested by \citet{Loeb1998} was to measure the frequency (or redshift) shift induced by the Hubble flow between multiple images of lensed sources, but the idea was rendered thus far impractical given the typical spectroscopic sensitivity, and, in part, given that the sources examined were generally extended, as well as the fact that the signal would be swamped by the frequency shift induced by the transverse peculiar motion of the lenses. Other innovative ideas \citep[e.g.][and references therein]{PiattellaGiani2017} were the possibility to measure the positional shift of multiple images of lensed sources, or the change in time delay (TD) between them, over a certain period (see also \citealt{BroadhurstOliver1991TD}). These ideas were also, mostly found to be currently impractical, give typical sensitivities and time periods required for their detection.

In this work we examine the option of probing such cosmological processes in ``real time" with lensed, repeating FRBs, and in particular, given the expected change in TD over time. In \S \ref{lensing} we outline a general lensing configuration and derive the TD expression in the thin lens approximation in a flat universe. We also derive the variations for a transverse moving source, for a source (and lens) drifting with the Hubble flow, as well as for a growing mass, and we outline explicit expressions for a point-mass lens, and for a singular isothermal sphere (SIS) lens. We numerically examine the effects  by comparing the TDs of various lensing configurations while changing the source (and lens) position, redshifts, and mass, verifying the results obtained using the analytic expectations. The results are presented in \S \ref{results}, and the work is concluded in \S \ref{conclusions}.

\section{ Gravitational Lensing}\label{lensing}

\begin{figure}
 \begin{center}
  \includegraphics[width=90mm,trim=4.5cm 6cm 3cm 5cm,clip]{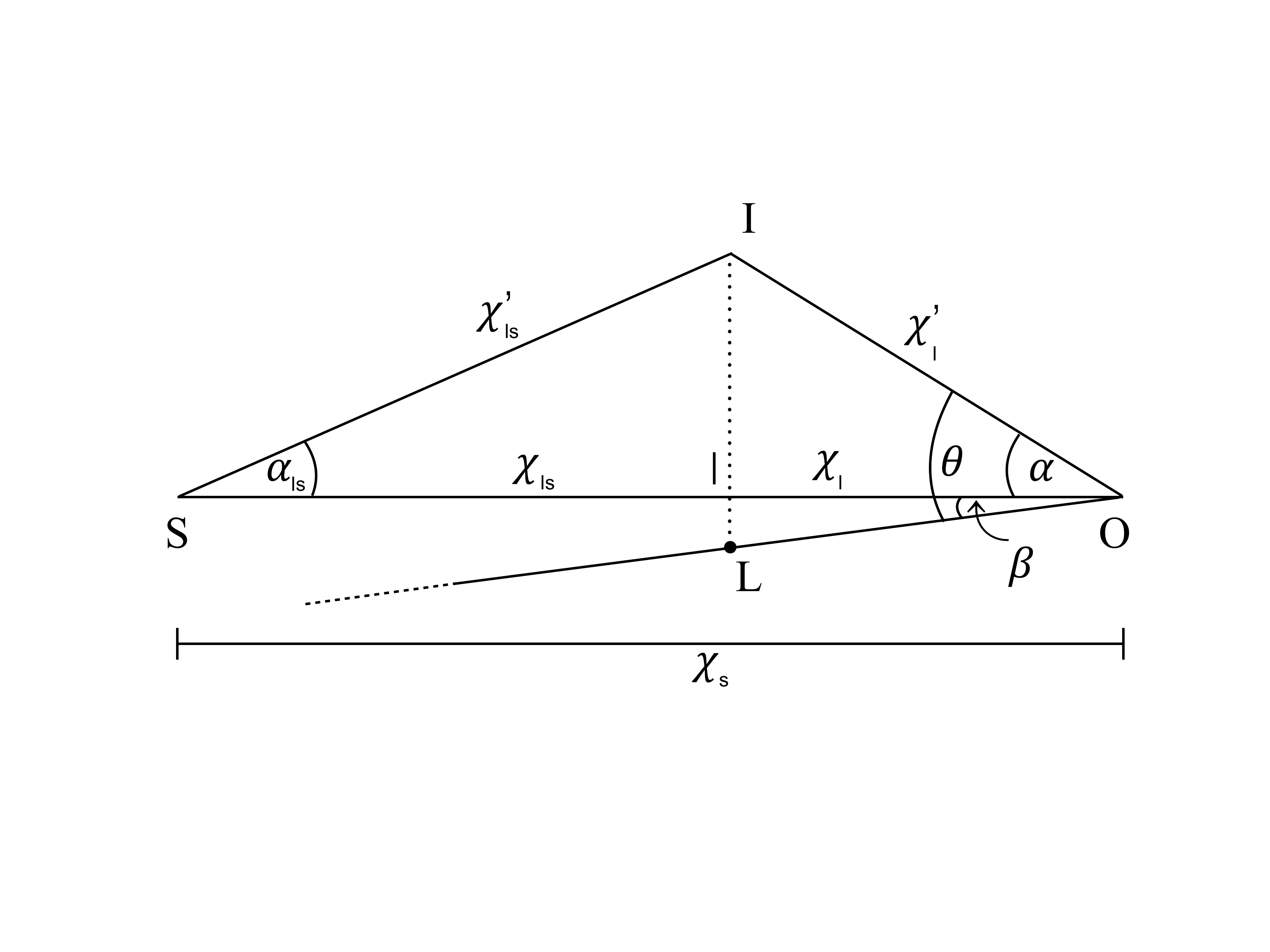}
 \end{center}
\caption{Illustration of a lensing configuration. The difference between the comoving deflected and undeflected path lengths results in the geometrical TD of the image. The difference in this delay between a pair of multiple images of the same source is the geometrical contribution to the observed TD.}\vspace{0.1cm}
\label{sketch}
\end{figure}

Assume a homogeneous, isotropic, zero curvature (k=0) universe.  Consider a source (S) nearly directly behind a gravitational lens (see Fig. \ref{sketch}). The source makes a real angle $\beta$ with the lens direction and let $\theta$ be the angle from the line of sight to the lens of its image (I), as seen by an observer (O).  Let $\chi_{l}$ be the (dimensionless) comoving radial distance to the lens plane which is defined to contain the lens and to be perpendicular to the line OS,  and denote point $l$ to be the intersection of the lens plane and the line OS.  Note that $l$ may vary with time, but if the source, lens and observer all follow the Hubble flow exactly, then $l$ also retains the same comoving coordinate. $\chi_{s}$ is the radial comoving distance between the observer and the source and $\chi_{ls}\equiv \chi_{s} -\chi_{l}$. 
The (reduced) angle of deflection $\alpha$ is $\theta-\beta$.
 
The dimensionless comoving distance to a redshift $z_{i}$, is defined as $\chi_{i}=\frac{c}{R_{o}}\int^{z_{i}}_{0}\frac{dz'}{H(z')}$, so that this comoving distance times the scale factor at the time the light reaches the observer, $R_{o}$, gives the proper physical distance to the object. The angular diameter distance as we shall use in some parts below is the proper distance times the scale factor at the time of emission, i.e., $D_{i}=\frac{c}{1+z_{i}}\int^{z_{i}}_{0}\frac{dz'}{H(z')}$. The angular diameter distances between two redshifts $z_{i}$ and $z_{j}$ is given by $D_{ij}=\frac{c}{1+z_{j}}\int^{z_{j}}_{z_{i}}\frac{dz'}{H(z')}$. 
The expansion rate of the Universe in the current, flat standard model (with $w=-1$, where $w$ equals the pressure over the energy density, i.e., the dark energy equation of state)  is,
\begin{equation}
H(z)=\sqrt{H_{0}^2(\Omega_{m,0}(1+z)^3+ \Omega_{\Lambda,0})} ,
\label{Hexpansion}
\end{equation}
with $H_{0}$ the present value of the Hubble constant.

The presence of the lens delays the arrival time of light rays from the source. The delay of an image due to the lens consists of two components
\begin{equation}
t= t_p +t_s,
\end{equation}
where $t_p$ is the geometrical TD due to the extra path length of the deflected ray, and $t_s$ is the delay due to the gravitational potential (known as the Shapiro delay; \citealt{Shapiro1964}).  
The geometrical term is given by noting that the (comoving) path length of the deflected ray, $\chi_{ls}'+\chi_{l}'$,  is longer than path length of the undeflected ray,  $\chi_{s}$, by:
\begin{eqnarray}
&c t_p&=R_o \left[ \chi_{ls}[cos\alpha_{ls}^{-1}-1]+\chi_l[cos\alpha^{-1}-1]\right],
\label{eq:tp}
\end{eqnarray}
where  $\tan \alpha_{ls} \equiv \chi_l \tan \alpha/\chi_{ls} $, and $R_o$ is the  scale factor of the universe at the time the light ray arrives at the observer. We do not normalize $R_o$ to unity as is often done because real-time cosmology needs to allow for the fact that $R_o$ changes with time. Rather the scale factor carries dimensions of length while comoving distances, as defined above, are dimensionless.
The gravitational potential contribution to the observed delay is given by:
\begin{equation}
t_s= \frac{-2(1+z_{l})}{c^3} \int \phi(s) ds
\end{equation}
where $\phi(s)$ is the gravitational potential along the path ds, and the $(1+z_{l})$ redshifts the Shapiro delay in the lens frame, $\frac{-2}{c^3}\int\phi(s)ds$, to the observer's point of view.

Note that equation (\ref{eq:tp}) reduces, in second order approximation in $\alpha$, to the geometrical part of the familiar time-delay formula (with, possibly, a slightly different definition of the distances; e.g., \citealt{Refsdal1964MNRAS, Schneider1985,BlandfordNarayan1986, Bartelmann2010reviewB}):
\begin{equation}\label{usualT}
t(\theta)=\frac{1+z_l}{c}\frac{D_{l}D_{s}}{D_{ls}} \left[ \frac{1}{2}(\theta-\beta)^2-\psi(\theta)  \right], 
\end{equation}
where $\theta$ is the observed image position and the effective lensing potential is $\psi(\theta)=\frac{2D_ls}{D_lD_sc^2}\int\phi ds$. In terms of comoving distances:
\begin{equation}\label{usualTComoving}
t(\theta)=\frac{R_o}{c}\frac{\chi_{l}\chi_{s}}{\chi_{ls}} \left[ \frac{1}{2}(\theta-\beta)^2-\psi(\theta)  \right], 
\end{equation}
with $\psi(\theta)=\frac{1+z_l}{R_0}\frac{2\chi_ls}{\chi_l\chi_sc^2}\int\phi ds$. Hereafter for our calculations we shall use these second-order formulae, although we stress that all results presented were also verified using higher orders of equation (\ref{eq:tp}) (instead of the full expression given in equation (\ref{eq:tp}); the series expansion of 1/cos(x) is often more stable numerically than explicitly employing 1/cos(x)). Note also that our formula is given in the thin lens approximation, and, that its derivation uses, as is customary (e.g. \citealt{Bartelmann2010reviewB}), the fact that the TD is very small compared to the Hubble time. 

The delay between two multiple images of the same background source, 1 and 2, observed at $\theta_1$ and $\theta_2$, is the difference in their respective delays $\Delta t \equiv t(\theta_2)-t(\theta_1)$.

\subsection{Lensing by a point mass}
The deflection angle at an angular distance $\theta$ from a point mass of mass $M$, located at a redshift of $z_{l}$, is given by \citep{NarayanBartelmann1996Lectures}:
\begin{equation}
\alpha(\theta)=\frac{4GM}{c^{2}\theta} \frac{D_{\rm ls}}{D_{\rm l}D_{\rm s}} , 
\label{def}
\end{equation}
where $D_{\rm l}$, $D_{\rm s}$, $D_{\rm ls}$ are the angular diameter distances to the lens, to the source (located at $z_{s}$), and between the lens and the source, respectively, $G$ the gravitational constant and $c$ the speed of light.  
The lensing potential for a point mass is given by:
\begin{equation}
\psi(\theta)=\frac{4GM}{c^{2}} \frac{D_{\rm ls}}{D_{\rm l}D_{\rm s}} \ln{|\theta |}.
\end{equation}\label{pot}
The Einstein radius of the lens is given by: 
\begin{equation}
  \theta_{\rm E} = \left(\frac{4GM}{c^2} \frac{D_{\rm ls}}{D_{\rm l}D_{\rm s}}\right)^{1/2} .
\label{re}
\end{equation}
The lens equation, describing the mapping between source and image planes, is generally given by:
\begin{equation}
  \beta(\theta) = \theta-\alpha(\theta), 
\label{lens_eq}
\end{equation}
where $\beta$ is the angular source position. For a point mass, this can be written as:
\begin{equation}
  \beta = \theta - \frac{\theta_{\rm E}^2}{\theta},
\label{lens_eq_point}
\end{equation}
which has two solutions -- so that two images of the same source appear, one on either side of the lens:
\begin{equation}
  \theta_\pm = \frac{1}{2}\,\left(\beta\pm\sqrt{\beta^{2}+4\theta_{\rm E}^{2}}\right).
\label{sols}
\end{equation}

The magnification of the two images, for a point mass lens, are given by:
\begin{equation}
  \mu_\pm = \left[1-\left(\frac{\theta_{\rm E}}{\theta_\pm}\right)^{4} \right]^{-1} = \frac{u^{2}+2}{2u\sqrt{u^{2}+4}}\pm\frac{1}{2} ,
\label{mags}
\end{equation}
where $u$ is the angular separation of the source from the point mass
in units of the Einstein angle, $u\equiv\beta\theta_{\rm E}^{-1}$.

The TD (following equation \ref{usualT}) between the two images expected for a source strongly lensed by a point mass, is  $\Delta t \equiv t(\theta_-) - t(\theta_+)$. Equivalently the (approximated) TD for a point lens can be written as \citep{Schneider1992}:
\begin{equation}
\Delta t=  \frac{4GM}{c^{3}} (1+z_{\rm l}) \tau(u) ,
\label{eqTDScheinder}
\end{equation}
where
\begin{equation}
  \tau(u)=\frac{1}{2}u\sqrt{u^{2}+4} + \ln\frac{\sqrt{u^{2}+4} +u}{\sqrt{u^{2}+4} -u}  .
\label{eqTDScheinder2}
\end{equation}

In what follows, we will examine the change in the TD $\Delta t$, between pairs of multiply imaged events (say, a and b), which is denoted (similar to \citealt{PiattellaGiani2017}):
\begin{equation}
  \Delta \equiv  \Delta t_{b}- \Delta t_{a}  .
\label{eqTDdif}
\end{equation}

\subsection{Lensing by a singular isothermal sphere (SIS)}
Perhaps a more realistic representation for a lens on galaxy scales, while maintaining a simple form, is the SIS. We shall thus compare our time delay results for the point lens also to this form and for completeness give here \citep{Schneider1992,Loeb1998} its potential:
\begin{equation}
\psi(\theta)=\frac{4G\sigma^2}{c^2} \frac{D_{\rm ls} }{D_{\rm s}} |\theta |,
\end{equation}\label{potSIS}

and particular time delay function:
\begin{equation}
\Delta t=\left( 4\pi\frac{\sigma^2}{c^2}\right)^2 \frac{D_lD_{ls}}{cD_{s}}(1+z_l)2u ,
\end{equation}
where and $\sigma$ is the 1D velocity dispersion and the Einstein radius is given by:
\begin{equation}
\theta_{E}=4\pi\frac{\sigma^2}{c^2}\frac{D_{ls}}{D_{s}}.
\end{equation}

\subsection{Time delay for a transverse moving source}

Consider the possibility that a repeating transient source sends two pulses a and b, that arrive in two images, 1 and 2,  per pulse. For simplicity, we consider the special case that the separation in time between the two pulses $t_b -t_a$, though of human timescale, is larger than the image delay $\Delta t$. The small change in the delay between two multiple images of the transversely moving source over a positional change of $\delta \beta$  is
 \begin{flalign} 
 \Delta &  = t_{b2} -t_{b1}  -t_{a2}  +t_{a1} = {{d t_2}\over{d \beta}}\delta \beta-{ {dt_1}\over{d\beta}} \delta \beta  && \\
 & =   \left[{\partial{ t_2}\over{\partial \theta}}|_{\beta}\frac{d\theta}{d\beta}+ {\partial{ t_2}\over{\partial \beta}}|_{\theta}   \nonumber
  -  {\partial{ t_1}\over{\partial \theta}}|_{\beta}\frac{d\theta}{d\beta}- {{\partial t_1}\over{\partial \beta}}|_{\theta}\right] \delta \beta && \\ \nonumber
& =\left[ ({\partial{ t_{2p}}\over{\partial \theta}}|_{\beta}  +    {\partial{ t_{2s}}\over{\partial \theta}}|_{\beta})\frac{d\theta}{d\beta}+ {\partial{ t_{2p}}\over{\partial \beta}}|_{\theta}  + {\partial{ t_{2s}}\over{\partial \beta}}|_{\theta}\right]\delta \beta \\ \nonumber
 & ~~~~ - \left[ ({\partial{ t_{1p}}\over{\partial \theta}}|_{\beta}  +    {\partial{ t_{1s}}\over{\partial \theta}}|_{\beta})\frac{d\theta}{d\beta}+ {\partial{ t_{1p}}\over{\partial \beta}}|_{\theta}  + {\partial{ t_{1s}}\over{\partial \beta}}|_{\theta}  \right] \delta \beta                                                      \nonumber                             
 \end{flalign}
where $\delta \beta = (d\beta/dt_0)(t_b -t_a)$. We denote the usual observed time as $t_0$ to distinguish it from the delayed time t.
  
Here  ${\partial{ t_{1,2}}\over{\partial \theta}}|_{\beta}$ means a virtual variation of $\theta$ keeping $\beta$ constant (even though the physical value of $\theta$ depends on $\beta$).  The quantities ${\partial{ t_{1s,2s}}\over{\partial \beta}}|_{\theta}$ vanish because the Shapiro delay $t_s$ depends only on $\theta$; there is no explicit dependence on $\beta$.

Now $t_p$ depends only on $\theta  -\beta$, so $\partial t_p/\partial \beta = -\partial t_p/\partial \theta $.
By Fermat's principle, which states that $t_p +t_s$ be a minimum over possible choices of $\theta $ for a given $\beta$, 
\begin{equation}
\partial t_p/\partial \theta  = -\partial t_s/\partial \theta 
\end{equation}
  and 
\begin{equation}
\partial t_p /\partial \beta = \partial t_s/\partial \theta .
\end{equation}
so finally

\begin{flalign}\label{TDchangeT}
\Delta & =\left[{\partial{ t_{2p}}\over{\partial \beta}}|_{\theta} -  {\partial{ t_{1p}}\over{\partial \beta}}|_{\theta}\right]\delta \beta .
\end{flalign}

Using the second-order TD equation (\ref{usualT}) one obtains:
\begin{equation}\label{TDchangeTPointMass}
\Delta =\frac{(1+z_{\rm l})}{c}\,
  \frac{D_{\rm l}D_{\rm s}}{D_{\rm ls}}\,
  \left[(\theta_{2}-\theta_{1})\right]\delta \beta .
\end{equation}
Alternatively, for a point mass, following equation (\ref{eqTDScheinder}) and using the chain rule this can be written as:
\begin{equation}
\Delta =\frac{4GM}{c^3}(1+z_{l})\frac{\sqrt{u^2+4}}{\theta_{E}} \delta \beta,
\end{equation}
and for a SIS lens, this can also be written simply as:
\begin{equation}\label{TDchangeTSIS}
\Delta =\Delta t \frac{\delta \beta}{\beta},
\end{equation}
so that $\frac{\delta \beta}{\beta}$ is the fractional change in angular source position.

\subsection{Time delay for a redshift-drifting source and lens}
There also exists the possibility of keeping $\beta$ fixed while changing the distances by a constant, which is expected just from the Hubble expansion itself. Consider now a similar, repeating transient source that sends two pulses a and b, that arrive in two images, 1 and 2, per pulse.  We consider the change that occurs just due to the Hubble flow and assume that the source, lens, and observer remain at constant comoving coordinates, hence the angle $\beta$ does not change.  The change in the delay  $ \Delta t $ is given by

\begin{equation}
\Delta  = t_{b2} -t_{b1}  -t_{a2}  +t_{a1} = {{d t_2}\over{d t_o}}\delta t_o-{{dt_1}\over{d t_o}} \delta  t_o .
\end{equation}

Note that a change in an image's arrival time,  ${d t_i}\over{d t_o}$, is given by:
\begin{flalign}
{{d t_i}\over{d t_o}} & ={\partial{ t_{i,p}}\over{\partial \theta}}|_{R_o}\frac{d\theta}{dt_o}+{\partial{ t_{i,p}}\over{\partial R_o}}|_{\theta} \frac{dR_o}{dt_o} \\
& + {\partial{ t_{i,s}}\over{\partial \theta}}|_{z_l}\frac{d\theta}{dt_o}+ {\partial{ t_{i,s}}\over{\partial z_l}}|_{\theta} \frac{dz_l}{dt_o} \nonumber .
\end{flalign}

Now note that the terms proportional to $\frac{d\theta}{dt_o}$ cancel out due to Fermat's principle, so we are left with:
 \begin{equation}
{{d t_i}\over{d t_o}}={\partial{ t_{i,p}}\over{\partial R_o}}|_{\theta} \frac{dR_o}{dt_o} + {\partial{ t_{i,s}}\over{\partial z_l}}|_{\theta} \frac{dz_l}{dt_o}.
\end{equation}

Recall that 
 \begin{equation}
\frac{dR_o}{dt}=R_o H_o,
\end{equation}
and
 \begin{equation}
\frac{dz_l}{dt_o}= (1+z_l)H_o - H(z_l), 
\end{equation}

so that 
 \begin{equation}
{{d t_i}\over{d t_o}}={\partial{ t_{i,p}}\over{\partial R_o}}|_{\theta} R_o H_o+ {\partial{ t_{i,s}}\over{\partial z_l}}|_{\theta} \left[(1+z_l)H_o - H(z_l)\right] .
\end{equation}

It is easy to see that
 \begin{equation}
{\partial{ t_{i,p}}\over{\partial R_o}}|_{\theta} R_o H_o = t_{i,p} H_o, 
\end{equation}
and 
 \begin{equation}
{\partial{ t_{i,s}}\over{\partial z_l}}=\frac{t_{i,s}}{(1+z_l)}. 
\end{equation}
Therefore
 \begin{eqnarray}
&{{d t_i}\over{d t_o}}=t_{i,p} H_o+\frac{t_{i,s}}{(1+z_l)} \left[(1+z_l)H_o - H(z_l)\right] \nonumber \\
&=t_iH_o-\frac{t_{i,s}}{(1+z_l)} H(z_l) .
\end{eqnarray}

Hence the overall change in TD between the two images, among the two events, is given by:
\begin{equation}\label{radialfinal}
\Delta =\left[\Delta t H_o-\frac{(t_{2,s}-t_{1,s})}{(1+z_l)} H(z_l)\right] \delta  t_o .
\end{equation}

where $\delta  t_o =(t_b - t_a)$, $t_b$ and $t_a$ being the average (cosmic) arrival times of pulses b and a at the observer.

In a similar manner we can derivate  the second-order TD equation (\ref{usualTComoving}) with respect to the observed time:
\begin{eqnarray}
&dt/dt_o=\frac{dR_o}{dt_o}\frac{1}{c}\frac{\chi_{l}\chi_{s}}{\chi_{ls}} \left[\frac{1}{2}(\theta-\beta)^2-\psi(\theta)  \right]   \nonumber \\
&+\frac{R_o}{c}\frac{\chi_{l}\chi_{s}}{\chi_{ls}} \frac{d}{dt_o}\left[\frac{1}{2}(\theta-\beta)^2-\psi(\theta)  \right] \nonumber \\
&=tH_o+\frac{R_o}{c}\frac{\chi_{l}\chi_{s}}{\chi_{ls}}\frac{d}{dt_o}\left[\frac{1}{2}(\theta-\beta)^2-\psi(\theta)  \right].
\end{eqnarray}

For a point mass, recalling Fermat's principle, and that $\psi(\theta)=\frac{4GM}{c^{2}} \frac{\chi_{\rm ls}}{\chi_{\rm l}\chi_{\rm s}} \frac{1+z_l}{R_o}\ln{(\theta)}$, it is easy to show that 
\begin{equation}
dt/dt_o=tH_o-\frac{R_o}{c}\frac{\chi_{\rm l}\chi_{\rm s}}{\chi_{\rm ls}}\psi(\theta)H(z_l)/(1+z_l),
\end{equation}
so that 
\begin{equation}\label{TDchangeR}
\Delta =\left[\Delta t H_o-\frac{R_o}{c}\frac{\chi_{\rm l}\chi_{\rm s}}{\chi_{\rm ls}}\left(\psi(\theta_{1})-\psi(\theta_{2})\right)\frac{H(z_l)}{(1+z_l)}\right] \delta t_o,
\end{equation}
in agreement with equation (\ref{radialfinal}). 

For a SIS lens, one obtains:
\begin{equation}\label{TDchangeRSIS}
\Delta =\left[\Delta t H_o + \frac{H(z_l)}{2(1+z_l)}\right] \delta t_o.
\end{equation}

\subsection{Time delay for a growing mass density}
The third option we examine is the effect of mass assembly on the TD. In other words we simply wish to measure the effect of a lens with increasing mass. As before we shall consider a repeating transient source that sends two pulses a and b, that arrive in two images, 1 and 2, per pulse. The small change in delay between the two multiple images of the source, due to a small increase in mass of the lens, is given by:
 \begin{equation} \label{mass}
 \Delta = {{d t_2}\over{d m}}\delta m-{{d t_1}\over{d m}}\delta m ,   
\end{equation}
where the $\delta m$ is simply $dm/dt_0 (t_b-t_a)$, and the exact expression for ${d t_i}\over{d m}$ depends on the lens type (or the actual mass distribution).  

For a point mass, differentiating equation (\ref{eqTDScheinder}) with respect to mass one obtains:
\begin{equation}
\Delta =\frac{4G}{c^3}(1+z_{l})\left[\tau(u) +M\frac{d\tau}{du}\frac{du}{d\theta_E}\frac{d\theta_E}{dM}\right] \delta M
\end{equation}

where
$\frac{d\tau}{du}=\sqrt{u^2+4}$; $\frac{du}{d\theta_E}=\frac{-\beta}{\theta_E^{2}}$; and $\frac{d\theta_E}{dM}=\frac{1}{2}\frac{\theta_E}{M}$, 

so that 
\begin{equation}\label{TDchangeM}
\Delta =\frac{4G}{c^3}(1+z_{l})\left[\tau(u) -\frac{u}{2}\sqrt{u^2+4}\right] \delta M .
\end{equation}

For a SIS lens one obtains:
\begin{equation}\label{TDchangeMSIS}
\Delta =\frac{\Delta t}{2} \frac{\delta M}{M} ,
\end{equation}
where $M$ in this case is the mass enclosed within the Einstein radius (equation (\ref{re})), and  $\frac{\delta M}{M}$ is simply the fractional change in this mass.
\\

\section{Results}\label{results}
We calculate the TD change, $\Delta$, over 1 year, for the three cases explored in the previous section: a transverse relative motion of the source and lens, a radial redshift drift with the Hubble flow, and a lens-mass increase, for different lens configurations (including different masses, redshifts, source positions, etc., for examples see Tables \ref{TablePointLens} and \ref{TableSISLens}). The results are calculated for a point mass using equations (\ref{TDchangeTPointMass}), (\ref{TDchangeR}), and (\ref{TDchangeM}), respectively, and for a SIS lens, using equations (\ref{TDchangeTSIS}), (\ref{TDchangeRSIS}), and (\ref{TDchangeMSIS}), respectively. We stress that all results are also verified numerically by simply calculating the TD, $\Delta t$, now, using equation (\ref{usualT}), and repeating the calculation for a year from now (i.e. after the relevant change is made), for the three cases. We also note that the choice of a year's time-span is arbitrary, and the results can be linearly scaled to any time period typical of the human lifetime. For the calculations we use the standard ${\Lambda\mathrm{CDM}}$ flat cosmological model with present-day Hubble constant, matter density and dark energy density, of $H_0 = 70$ $  \mathrm{km~s^{-1}~Mpc^{-1}}$, $\mathrm{\Omega_{m,0}}=0.3$, and $\mathrm{\Omega_{\Lambda,0}}=0.7$, respectively.

\subsection{Transverse displacement}\label{s:transverse}
We shall now examine a test case for a transverse displacement of the source compared to the observer-lens line-of sight. Note that in practice the source is the one we assume is moving with respect to the observer-lens line of sight, which entails only a small shift in the source position $\beta$. Note also that as we only aim to probe here the effect of the displacement of the source, we do not take into account the fact that the universe is slightly larger for the second event; an effect we will probe separately below (\S \ref{s:radial}). We thus check now how a slight transverse source displacement propagates into the TD, using test-case lensing configurations with a point-mass lens and a SIS lens, where the source moves transversely between, say, two repeating FRB events (let them be called a and b), each event being multiply imaged.

We also note that if the lens changes (i.e., moves) while the light travels from the source to us, this may additionally cause a slight frequency shift between the two images of either event. The frequency (or redshift) difference between two multiple images is of order $\beta_T\alpha$, with $\alpha$ the deflection angle and $\beta_T=v_{T}/c$ where $v_{T}$ is the transverse velocity of the lens relative to the observer-source line-of-sight. As discussed in previous work \citep[][see also \citealt{BirkinshawGull1983,Frittelli2003,Sereno2008,Molnar2013}]{Loeb1998}, for images of a source lensed by a galaxy (galaxy cluster) moving transversely with typical cosmic velocities, this frequency shift can translate into a measured radial velocity shift of order a few (few dozen) m/s between the multiple images.

Assume that the relative transverse velocity of the source in a rest-frame defined with respect to the observer-lens line-of-sight (LOS), is 500 km/s, so that in one year in the source plane it moves a proper distance of $1.58 \times 10^{10}$ km. For a source at $z_{s}=2$ each arcsecond corresponds to $\simeq2.58 \times 10^{17}$ km, so that the source moves in \emph{an observed} year, $\delta \beta = 0.61\times10^{-7}/(1+z_{s})= 0.2\times10^{-7}$ seconds of an arc with respect to the LOS to the lens. Therefore, to examine the change in the respective TDs of the two events (that is, to compare the TD between the two images of the first event, event a, with the TD between the two images of the second event, event b), we simply (numerically) calculate the above equation (\ref{usualT}), once for a source at a certain angular distance $\beta$ from the LOS, corresponding to event a, and then for a source at  $\beta +\delta \beta$ from the LOS, which is event b. The difference corresponds to $\Delta $ (equation (\ref{eqTDdif})), and matches that given by equation (\ref{TDchangeTPointMass}) and (\ref{TDchangeTSIS}) (the analytical expectations). We repeat the calculation and obtain $\Delta$ for various initial values of lens mass, source position, lens and source redshifts, etc. We obtain, for example, $\Delta\sim0.6$ sec for a $10^{12} M_{\odot}$ point mass lens, with $z_{l}=0.5$, $z_{s}=2.0$,  and for various values of $\beta$. For a more massive lens, $10^{13} M_{\odot}$, we obtain for various source positions differences of $\Delta\sim2$ sec between TDs of the two events. More lens configurations are seen in Tables \ref{TablePointLens} and \ref{TableSISLens}, including for a SIS lens. We conclude that the transverse motion causes a TD change of order a few seconds per year.

\subsection{Redshift drift}\label{s:radial}
Here again we assume two events multiply imaged each, a and b, separated by a certain time. We start with a lens at $z_{l}=0.5$ and $z_{s}=2.0$. With our said choice of cosmological parameters, the TD between the two images of event b, observed a year later than event a, is larger by $\simeq1\times10^{-4}$ sec than the TD between the two images of event a  -- for a small point mass of $10^{11} M_{\odot}$ and a source position of $\beta=0.3"$ (TD of about 32 days). Using a two-times larger angular source distance from the LOS increases the TD, and the TD difference, by about a factor of two, so that in that case $TD\sim66$ days, and $\Delta \simeq2.4\times10^{-4}$ seconds. Increasing the mass of the lens to $10^{12} M_{\odot}$, for example, yields TDs of about a year, and  - for observed events separated by, say, 5 years, $\Delta\simeq0.004$ sec if $\beta=0.6"$, or $\Delta \simeq0.01$ sec for $\beta=1.5"$. A SIS lens with similar masses yields, as expected, similar results (see Tables \ref{TablePointLens} and \ref{TableSISLens}). So the yearly TD change due to the redshift drift is, for most practical cases, of order $\Delta \sim 10^{-2}-10^{-4}$ sec.  This small change is detectable with FRBs, but could be masked by the much larger delay due to transverse motion, unless the larger effect, which is equally likely to be positive or negative, is somehow isolated and removed from the smaller effect, which always has the same sign.  As the ratio of the two effects is of order $10^3$, this would be challenging to say the least.

\subsection{Mass assembly}\label{s:mass}
As structures collapse more mass is being accreted or moves towards the center of the lens. This means that the lens mass effectively increases with time and we set to examine the effect on the measured TD in a similar fashion to the above. Given that a year constitutes about $\sim10^{-10}$ of the Hubble time, we aim to test what TD shift is caused by a mass increase of that order. We find in this case that for a $10^{12} M_{\odot}$ point lens, with $z_{l}=0.5$, $z_{s}=2.0$ and $\beta=1"$ (TD of about a year), the TD change is $ \Delta \simeq 0.0014$ sec. For a $\beta=0.3"$ (TD of about three months), for example, the TD change becomes $\simeq 4\times10^{-4}$ sec. Increasing the mass to $10^{13} M_{\odot}$, with $\beta=0.15"$ (TD of almost six months), $\Delta \simeq 7\times10^{-4}$ sec, and for $\beta=1"$ (TD of 3 years), $\Delta \simeq0.005$ sec per year.  Tables (\ref{TablePointLens} \& \ref{TableSISLens}) list few other examples for both point mass and SIS lens. The TD increases with lens mass and and source position so we conclude that overall, the yearly TD change due to growth of perturbations or mass assembly, is typically of order $\Delta \sim10^{-4}-10^{-2}$ for $\sim10^{11}- \sim10^{13}$ lenses for relevant source positions.

\section{Conclusions}\label{conclusions}
We have analytically and numerically examined the expected change in TD between two multiple images of a persisting or repeating lensed source across time. Using point-mass and SIS lenses, we have found that in as short a period as 1 year, a TD change of about a second (or $\sim 10^{-1}-10^{1}$ s), is expected for a transverse moving source with typical transverse velocities and lensed with typical galaxy or cluster masses; a TD change of $\sim10^{-4} -10^{-2}$ s is expected over this year just from the redshift drift with the Hubble flow; and a TD change of $\sim10^{-4} -10^{-2}$ s is expected from mass assembly as structures build up. The magnitude of all these effects is measurable. By accurately measuring the arrival time of multiply imaged, repeating FRB, real-time cosmology with FRBs could thus soon be feasible.

The most dominant effect is thus the TD change caused by a relative transverse motion between the observer, lens and source (which we simplified in practice to a source displacement). The sign of the TD change depends on the relative transverse velocity direction: if the source approaches the lens on the sky then the TD between multiple images will continuously decrease, whilst as the source moves away from the lens the TD will increase. Thus, if one were to measure the TD change over a course of some finite time, given the amplitude of the effect compared to the other two probed here, this will directly enable a measurement of the relative transverse velocity of the objects involved. In other words real-time measurement of transverse velocities could readily be done, if and when strongly lensed repeating FRBs are found.

One way to (partially) remove the effect of transverse velocities overcasting the redshift-drift effect, may be measuring the frequency shift expected due to the relative transverse velocity of the lens. In addition, statistically, since the source-lens relative displacement should be equal in all directions, this effect should average out to zero over a very large sample of sources.  In practice, though, the transverse velocities are likely to be the dominant effect in the near future.  Measuring the transverse velocities as a function of redshift, however, could be an important test of theories of cosmic structure formation. 

We do not discuss here effects due to microlensing or other structural changes within the lenses, which are expected to affect the TDs \citep[e.g.][]{TieKochanek2018,Goldstein2018}; however, in fact, microlensing in the lensing galaxy has long been proposed as a means to measure the transverse velocity of the lensing galaxy relative to the observer-source line of sight \citep{Grieger1986TransverseMicrolensing,Gould1995TransverseMicorlensing,Mediavilla2016TravsverseMicrolensing}, and thus one can imagine such measurements could aid in recovering the underlying redshift-drift signal.

Lastly we comment on the ideal lenses used. In reality lenses will not be circular but approximately elliptical (for example, for galaxies) or more complex (e.g., for galaxy groups and clusters). For non-singular lenses, more than two images of a background source will often appear (especially if non-circular), which, while these might slightly complicate the interpretation in cases where there is no sufficient spatial resolution, they will also allow for more TD measurements per lensed source, including with shorter wait times (which would be important for some of the larger lenses).

We conclude that real-time processes in the universe might already be measured with current radio observation time resolution, and particularly, using repeating, lensed FRBs, which easily allow time resolution down to 1 millisecond, and possibly  even better \citep[e.g.,][]{Zheng2014FRBs,Eichler2017FRBs}. The dominant effect, led by the relative motion of the lens and the source, is readily measurable with current instruments, while the smaller (but no less important) effects of redshift-drift and mass build-up will have to be distinguished  statistically or by new means. Potentially, other persisting or repeating sources, such as distant quasars, lensed gamma-ray bursts with radio counterparts (\citealt{Barnacka2016GRB} and references therein), or other speculative examples such as cosmological pulsars \citep{Kim2015RealTimeCosmology}, might also become useful sometime in the near future. \\
\\

We thank J. Wagner for helpful discussions. Some of our distance calculations rely on useful scripts by \citet{Ofek2014Matlab}. We acknowledge the support of the Israel Science Foundation, the Israel-U.S. Binational Science Foundation, and the Joan and Robert Arnow Chair of Theoretical Astrophysics for financial support of this research. 

\begin{deluxetable*}{ccccl||lllc}
\tablecaption{\small{Time delay and its yearly change for different lensing configurations for a point mass} \label{TablePointLens}}
\tablehead{
\colhead{$z_{lens}$
} &
\colhead{$z_{source}$
} &
\colhead{$\beta$ 
} &
\colhead{M [$M_{\odot}$]
}  &
\colhead{$\theta_E$ ["]
}  &
\colhead{$\Delta$ t~ [d]
} &
\colhead{$\Delta_{T}$ [s]
}
&
\colhead{$\Delta_{R}$ [s]
}
&
\colhead{$\Delta_{M}$ [s]
}}
\startdata
0.20 & 2.00 & 0.36 & 3.2e+11 & 1.79 & 34.7 & 0.17 & 1.16e-04 & 1.49e-04 \\ 
0.20 & 2.00 & 0.89 & 3.2e+11 & 1.79 & 87.49 & 0.18 & 2.97e-04 & 3.70e-04 \\ 
0.20 & 2.00 & 1.13 & 3.2e+12 & 5.65 &  347 & 0.54 & 1.16e-03 & 1.49e-03 \\ 
0.20 & 2.00 & 2.82 & 3.2e+12 & 5.65 & 874.9 & 0.56 & 2.97e-03 & 3.70e-03 \\ 
0.20 & 2.00 & 6.35 & 1.0e+14 & 31.76 & 1.097e+04 & 3.05 & 3.67e-02 & 4.72e-02 \\ 
0.20 & 2.00 & 15.88 & 1.0e+14 & 31.76 & 2.767e+04 & 3.13 & 9.40e-02 & 1.17e-01 \\ 
0.20 & 5.00 & 0.37 & 3.2e+11 & 1.84 & 34.7 & 0.11 & 1.16e-04 & 1.49e-04 \\ 
0.20 & 5.00 & 0.92 & 3.2e+11 & 1.84 & 87.49 & 0.11 & 2.97e-04 & 3.70e-04 \\ 
0.20 & 5.00 & 1.16 & 3.2e+12 & 5.82 &  347 & 0.35 & 1.16e-03 & 1.49e-03 \\ 
0.20 & 5.00 & 2.91 & 3.2e+12 & 5.82 & 874.9 & 0.36 & 2.97e-03 & 3.70e-03 \\ 
0.20 & 5.00 & 6.55 & 1.0e+14 & 32.74 & 1.097e+04 & 1.97 & 3.67e-02 & 4.72e-02 \\ 
0.20 & 5.00 & 16.37 & 1.0e+14 & 32.74 & 2.767e+04 & 2.02 & 9.39e-02 & 1.17e-01 \\ 
0.50 & 2.00 & 0.23 & 3.2e+11 & 1.14 & 43.37 & 0.34 & 1.52e-04 & 1.87e-04 \\ 
0.50 & 2.00 & 0.57 & 3.2e+11 & 1.14 & 109.4 & 0.34 & 3.87e-04 & 4.63e-04 \\ 
0.50 & 2.00 & 0.72 & 3.2e+12 & 3.61 & 433.7 & 1.06 & 1.52e-03 & 1.87e-03 \\ 
0.50 & 2.00 & 1.80 & 3.2e+12 & 3.61 & 1094 & 1.09 & 3.87e-03 & 4.63e-03 \\ 
0.50 & 2.00 & 4.06 & 1.0e+14 & 20.28 & 1.372e+04 & 5.96 & 4.79e-02 & 5.90e-02 \\ 
0.50 & 2.00 & 10.14 & 1.0e+14 & 20.28 & 3.458e+04 & 6.12 & 1.22e-01 & 1.46e-01 \\ 
0.50 & 5.00 & 0.25 & 3.2e+11 & 1.24 & 43.37 & 0.20 & 1.52e-04 & 1.87e-04 \\ 
0.50 & 5.00 & 0.62 & 3.2e+11 & 1.24 & 109.4 & 0.21 & 3.87e-04 & 4.63e-04 \\ 
0.50 & 5.00 & 0.79 & 3.2e+12 & 3.94 & 433.7 & 0.65 & 1.52e-03 & 1.87e-03 \\ 
0.50 & 5.00 & 1.97 & 3.2e+12 & 3.94 & 1094 & 0.66 & 3.87e-03 & 4.63e-03 \\ 
0.50 & 5.00 & 4.43 & 1.0e+14 & 22.14 & 1.372e+04 & 3.64 & 4.79e-02 & 5.91e-02 \\ 
0.50 & 5.00 & 11.07 & 1.0e+14 & 22.14 & 3.458e+04 & 3.73 & 1.22e-01 & 1.46e-01 \\ 
1.00 & 2.00 & 0.15 & 3.2e+11 & 0.75 & 57.83 & 0.68 & 2.01e-04 & 2.49e-04 \\ 
1.00 & 2.00 & 0.38 & 3.2e+11 & 0.75 & 145.8 & 0.70 & 5.13e-04 & 6.17e-04 \\ 
1.00 & 2.00 & 0.48 & 3.2e+12 & 2.38 & 578.3 &2.15 & 2.01e-03 & 2.49e-03 \\ 
1.00 & 2.00 & 1.19 & 3.2e+12 & 2.38 & 1458 & 2.20 & 5.13e-03 & 6.17e-03 \\ 
1.00 & 2.00 & 2.67 & 1.0e+14 & 13.37 & 1.829e+04 & 12.07 & 6.34e-02 & 7.87e-02 \\ 
1.00 & 2.00 & 6.68 & 1.0e+14 & 13.37 & 4.611e+04 & 12.38 & 1.62e-01 & 1.95e-01 \\ 
1.00 & 5.00 & 0.19 & 3.2e+11 & 0.95 & 57.83 & 0.36 & 2.01e-04 & 2.49e-04 \\ 
1.00 & 5.00 & 0.47 & 3.2e+11 & 0.95 & 145.8 & 0.37 & 5.13e-04 & 6.17e-04 \\ 
1.00 & 5.00 & 0.60 & 3.2e+12 & 3.00 & 578.3 & 1.13 & 2.01e-03 & 2.49e-03 \\ 
1.00 & 5.00 & 1.50 & 3.2e+12 & 3.00 & 1458 & 1.16 & 5.13e-03 & 6.17e-03 \\ 
1.00 & 5.00 & 3.37 & 1.0e+14 & 16.84 & 1.829e+04 & 6.38 & 6.34e-02 & 7.87e-02 \\ 
1.00 & 5.00 & 8.42 & 1.0e+14 & 16.84 & 4.611e+04 & 6.54 & 1.62e-01 & 1.95e-01 \\ 
2.00 & 8.00 & 0.16 & 3.2e+11 & 0.79 & 86.74 & 0.56 & 2.72e-04 & 3.74e-04 \\ 
2.00 & 8.00 & 0.40 & 3.2e+11 & 0.79 & 218.7 & 0.57 & 6.97e-04 & 9.26e-04 \\ 
2.00 & 8.00 & 0.50 & 3.2e+12 & 2.51 & 867.4 & 1.77 & 2.72e-03 & 3.73e-03 \\ 
2.00 & 8.00 & 1.25 & 3.2e+12 & 2.51 & 2187 & 1.81 & 6.97e-03 & 9.26e-03 \\ 
2.00 & 8.00 & 2.82 & 1.0e+14 & 14.09 & 2.743e+04 & 9.94 & 8.60e-02 & 1.18e-01 \\ 
2.00 & 8.00 & 7.05 & 1.0e+14 & 14.09 & 6.917e+04 & 10.20 & 2.20e-01 & 2.93e-01 \\ 
2.00 & 10.00 & 0.16 & 3.2e+11 & 0.82 & 86.74 & 0.51 & 2.72e-04 & 3.73e-04 \\ 
2.00 & 10.00 & 0.41 & 3.2e+11 & 0.82 & 218.7 & 0.52 & 6.97e-04 & 9.26e-04 \\ 
2.00 & 10.00 & 0.52 & 3.2e+12 & 2.60 & 867.4 & 1.62 & 2.72e-03 & 3.73e-03 \\ 
2.00 & 10.00 & 1.30 & 3.2e+12 & 2.60 & 2187 & 1.66 & 6.97e-03 & 9.26e-03 \\ 
2.00 & 10.00 & 2.92 & 1.0e+14 & 14.60 & 2.743e+04 & 9.10 & 8.60e-02 & 1.18e-01 \\ 
2.00 & 10.00 & 7.30 & 1.0e+14 & 14.60 & 6.917e+04 & 9.33 & 2.20e-01 & 2.93e-01 \\ 
5.00 & 8.00 & 0.10 & 3.2e+11 & 0.51 & 173.5 & 1.74 & 3.51e-04 & 7.47e-04 \\ 
5.00 & 8.00 & 0.25 & 3.2e+11 & 0.51 & 437.5 & 1.78 & 9.17e-04 & 1.85e-03 \\ 
5.00 & 8.00 & 0.32 & 3.2e+12 & 1.61 & 1735 & 5.50 & 3.51e-03 & 7.47e-03 \\ 
5.00 & 8.00 & 0.81 & 3.2e+12 & 1.61 & 4375 & 5.64 & 9.17e-03 & 1.85e-02 \\ 
5.00 & 8.00 & 1.81 & 1.0e+14 & 9.06 & 5.486e+04 & 30.92 & 1.11e-01 & 2.36e-01 \\ 
5.00 & 8.00 & 4.53 & 1.0e+14 & 9.06 & 1.383e+05 & 31.72 & 2.90e-01 & 5.86e-01 \\ 
5.00 & 10.00 & 0.12 & 3.2e+11 & 0.59 & 173.5 & 1.42 & 3.51e-04 & 7.47e-04 \\ 
5.00 & 10.00 & 0.30 & 3.2e+11 & 0.59 & 437.5 & 1.45 & 9.17e-04 & 1.85e-03 \\ 
5.00 & 10.00 & 0.37 & 3.2e+12 & 1.87 & 1735 & 4.48 & 3.51e-03 & 7.47e-03 \\ 
5.00 & 10.00 & 0.94 & 3.2e+12 & 1.87 & 4375 & 4.60 & 9.17e-03 & 1.85e-02 \\ 
5.00 & 10.00 & 2.11 & 1.0e+14 & 10.53 & 5.486e+04 & 25.21 & 1.11e-01 & 2.36e-01 \\ 
5.00 & 10.00 & 5.27 & 1.0e+14 & 10.53 & 1.383e+05 & 25.86 & 2.90e-01 & 5.86e-01
\enddata
\tablecomments{The left-hand-side part of the Table are the input parameters, where the right-hand-side are the output time delay properties.\\
$\emph{Column 1:}$ Lens redshift.\\
$\emph{Column 2:}$ Source redshift.\\
$\emph{Column 3:}$ Source angular position, in arcseconds.\\
$\emph{Column 4:}$ Mass of lens.\\
$\emph{Column 5:}$ Einstein radius, in arcseconds.\\
$\emph{Column 6:}$ Time delay, in days. \\
$\emph{Column 7:}$ Change in the time delay over a year's time from transverse motion (see text for more details).\\
$\emph{Column 8:}$ Change in the time delay over a year's time from redshift drift (see text for more details).\\
$\emph{Column 9:}$ Change in the time delay over a year's time from mass increase (see text for more details).\\
}
\end{deluxetable*}
 \newpage
 
\begin{deluxetable*}{cccccl||lllc}
\tablecaption{\small{Time delay and its yearly change for different lensing configurations for a SIS lens} \label{TableSISLens}}
\tablehead{
\colhead{$z_{lens}$
} &
\colhead{$z_{source}$
} &
\colhead{$\beta$ 
} &
\colhead{$\sigma$ [km/s]
} & 
\colhead{$\theta_E$ ["]
} &
\colhead{$M_E$ [$M_{\odot}$]
} &
\colhead{$\Delta$ t~ [d]
} &
\colhead{$\Delta_{T}$ [s]
}
&
\colhead{$\Delta_{R}$ [s]
}
&
\colhead{$\Delta_{M}$ [s]
}}
\startdata
0.20 & 2.00 & 0.44 & 300 & 2.19 &  3.4e+11 &  51.91 & 0.21 & 3.21e-04 & 2.24e-04 \\ 
0.20 & 2.00 & 1.09 & 300 & 2.19 &  3.4e+11 &  129.8 & 0.21 & 8.02e-04 & 5.61e-04 \\ 
0.20 & 2.00 & 1.75 & 600 & 8.75 &  5.4e+12 &  830.6 & 0.83 & 5.13e-03 & 3.59e-03 \\ 
0.20 & 2.00 & 4.37 & 600 & 8.75 &  5.4e+12 &  2076 & 0.83 & 1.28e-02 & 8.97e-03 \\ 
0.20 & 2.00 & 4.86 & 1000 & 24.29 &  4.2e+13 &  6409 & 2.32 & 3.96e-02 & 2.77e-02 \\ 
0.20 & 2.00 & 12.15 & 1000 & 24.29 &  4.2e+13 &  1.602e+04 & 2.32 & 9.90e-02 & 6.92e-02 \\ 
0.20 & 5.00 & 0.46 & 300 & 2.32 &  4.0e+11 &  55.15 & 0.14 & 3.41e-04 & 2.38e-04 \\ 
0.20 & 5.00 & 1.16 & 300 & 2.32 &  4.0e+11 &  137.9 & 0.14 & 8.52e-04 & 5.96e-04 \\ 
0.20 & 5.00 & 1.86 & 600 & 9.29 &  6.5e+12 &  882.5 & 0.56 & 5.45e-03 & 3.81e-03 \\ 
0.20 & 5.00 & 4.65 & 600 & 9.29 &  6.5e+12 &  2206 & 0.56 & 1.36e-02 & 9.53e-03 \\ 
0.20 & 5.00 & 5.16 & 1000 & 25.81 &  5.0e+13 &  6809 & 1.54 & 4.21e-02 & 2.94e-02 \\ 
0.20 & 5.00 & 12.91 & 1000 & 25.81 &  5.0e+13 &  1.702e+04 & 1.54 & 1.05e-01 & 7.35e-02 \\ 
0.50 & 2.00 & 0.33 & 300 & 1.65 &  2.7e+11 &  90.55 & 0.48 & 5.60e-04 & 3.91e-04 \\ 
0.50 & 2.00 & 0.82 & 300 & 1.65 &  2.7e+11 &  226.4 & 0.48 & 1.40e-03 & 9.78e-04 \\ 
0.50 & 2.00 & 1.32 & 600 & 6.60 &  4.3e+12 &  1449 & 1.93 & 8.96e-03 & 6.26e-03 \\ 
0.50 & 2.00 & 3.30 & 600 & 6.60 &  4.3e+12 &  3622 & 1.93 & 2.24e-02 & 1.56e-02 \\ 
0.50 & 2.00 & 3.66 & 1000 & 18.32 &  3.3e+13 &  1.118e+04 & 5.36 & 6.91e-02 & 4.83e-02 \\ 
0.50 & 2.00 & 9.16 & 1000 & 18.32 &  3.3e+13 &  2.795e+04 & 5.36 & 1.73e-01 & 1.21e-01 \\ 
0.50 & 5.00 & 0.39 & 300 & 1.97 &  4.5e+11 &  107.9 & 0.32 & 6.67e-04 & 4.66e-04 \\ 
0.50 & 5.00 & 0.98 & 300 & 1.97 &  4.5e+11 &  269.7 & 0.32 & 1.67e-03 & 1.17e-03 \\ 
0.50 & 5.00 & 1.57 & 600 & 7.86 &  7.2e+12 &  1726 & 1.29 & 1.07e-02 & 7.46e-03 \\ 
0.50 & 5.00 & 3.93 & 600 & 7.86 &  7.2e+12 &  4316 & 1.29 & 2.67e-02 & 1.86e-02 \\ 
0.50 & 5.00 & 4.37 & 1000 & 21.83 &  5.6e+13 &  1.332e+04 & 3.57 & 8.23e-02 & 5.75e-02 \\ 
0.50 & 5.00 & 10.92 & 1000 & 21.83 &  5.6e+13 &  3.33e+04 & 3.57 & 2.06e-01 & 1.44e-01 \\ 
1.00 & 2.00 & 0.19 & 300 & 0.94 &  6.5e+10 &  90.29 & 0.84 & 5.58e-04 & 3.90e-04 \\ 
1.00 & 2.00 & 0.47 & 300 & 0.94 &  6.5e+10 &  225.7 & 0.84 & 1.40e-03 & 9.75e-04 \\ 
1.00 & 2.00 & 0.75 & 600 & 3.76 &  1.0e+12 &  1445 & 3.38 & 8.93e-03 & 6.24e-03 \\ 
1.00 & 2.00 & 1.88 & 600 & 3.76 &  1.0e+12 &  3612 & 3.38 & 2.23e-02 & 1.56e-02 \\ 
1.00 & 2.00 & 2.09 & 1000 & 10.45 &  8.0e+12 &  1.115e+04 & 9.38 & 6.89e-02 & 4.82e-02 \\ 
1.00 & 2.00 & 5.22 & 1000 & 10.45 &  8.0e+12 &  2.787e+04 & 9.38 & 1.72e-01 & 1.20e-01 \\ 
1.00 & 5.00 & 0.30 & 300 & 1.49 &  2.6e+11 &  143.4 & 0.56 & 8.86e-04 & 6.19e-04 \\ 
1.00 & 5.00 & 0.75 & 300 & 1.49 &  2.6e+11 &  358.4 & 0.56 & 2.22e-03 & 1.55e-03 \\ 
1.00 & 5.00 & 1.19 & 600 & 5.97 &  4.2e+12 &  2294 & 2.25 & 1.42e-02 & 9.91e-03 \\ 
1.00 & 5.00 & 2.99 & 600 & 5.97 &  4.2e+12 &  5735 & 2.25 & 3.54e-02 & 2.48e-02 \\ 
1.00 & 5.00 & 3.32 & 1000 & 16.59 &  3.2e+13 &  1.77e+04 & 6.25 & 1.09e-01 & 7.65e-02 \\ 
1.00 & 5.00 & 8.29 & 1000 & 16.59 &  3.2e+13 &  4.425e+04 & 6.25 & 2.74e-01 & 1.91e-01 \\ 
2.00 & 8.00 & 0.22 & 300 & 1.09 &  1.1e+11 &  164.5 & 0.77 & 1.02e-03 & 7.11e-04 \\ 
2.00 & 8.00 & 0.55 & 300 & 1.09 &  1.1e+11 &  411.2 & 0.77 & 2.54e-03 & 1.78e-03 \\ 
2.00 & 8.00 & 0.87 & 600 & 4.37 &  1.7e+12 &  2632 & 3.07 & 1.63e-02 & 1.14e-02 \\ 
2.00 & 8.00 & 2.18 & 600 & 4.37 &  1.7e+12 &  6579 & 3.07 & 4.07e-02 & 2.84e-02 \\ 
2.00 & 8.00 & 2.43 & 1000 & 12.14 &  1.3e+13 &  2.031e+04 & 8.52 & 1.26e-01 & 8.77e-02 \\ 
2.00 & 8.00 & 6.07 & 1000 & 12.14 &  1.3e+13 &  5.076e+04 & 8.52 & 3.14e-01 & 2.19e-01 \\ 
2.00 & 10.00 & 0.23 & 300 & 1.17 &  1.3e+11 &  176.4 & 0.73 & 1.09e-03 & 7.62e-04 \\ 
2.00 & 10.00 & 0.59 & 300 & 1.17 &  1.3e+11 &   441 & 0.73 & 2.73e-03 & 1.91e-03 \\ 
2.00 & 10.00 & 0.94 & 600 & 4.69 &  2.1e+12 &  2822 & 2.91 & 1.74e-02 & 1.22e-02 \\ 
2.00 & 10.00 & 2.34 & 600 & 4.69 &  2.1e+12 &  7056 & 2.91 & 4.36e-02 & 3.05e-02 \\ 
2.00 & 10.00 & 2.60 & 1000 & 13.02 &  1.6e+13 &  2.178e+04 & 8.07 & 1.35e-01 & 9.41e-02 \\ 
2.00 & 10.00 & 6.51 & 1000 & 13.02 &  1.6e+13 &  5.444e+04 & 8.07 & 3.37e-01 & 2.35e-01 \\ 
5.00 & 8.00 & 0.07 & 300 & 0.34 &  2.4e+09 &  76.61 & 1.15 & 4.74e-04 & 3.31e-04 \\ 
5.00 & 8.00 & 0.17 & 300 & 0.34 &  2.4e+09 &  191.5 & 1.15 & 1.18e-03 & 8.27e-04 \\ 
5.00 & 8.00 & 0.27 & 600 & 1.36 &  3.8e+10 &  1226 & 4.60 & 7.58e-03 & 5.30e-03 \\ 
5.00 & 8.00 & 0.68 & 600 & 1.36 &  3.8e+10 &  3065 & 4.60 & 1.89e-02 & 1.32e-02 \\ 
5.00 & 8.00 & 0.75 & 1000 & 3.77 &  2.9e+11 &  9459 & 12.79 & 5.85e-02 & 4.09e-02 \\ 
5.00 & 8.00 & 1.88 & 1000 & 3.77 &  2.9e+11 &  2.365e+04 & 12.79 & 1.46e-01 & 1.02e-01 \\ 
5.00 & 10.00 & 0.09 & 300 & 0.46 &  5.9e+09 &  103.5 & 1.09 & 6.40e-04 & 4.47e-04 \\ 
5.00 & 10.00 & 0.23 & 300 & 0.46 &  5.9e+09 &  258.7 & 1.09 & 1.60e-03 & 1.12e-03 \\ 
5.00 & 10.00 & 0.37 & 600 & 1.83 &  9.4e+10 &  1656 & 4.36 & 1.02e-02 & 7.15e-03 \\ 
5.00 & 10.00 & 0.92 & 600 & 1.83 &  9.4e+10 &  4139 & 4.36 & 2.56e-02 & 1.79e-02 \\ 
5.00 & 10.00 & 1.02 & 1000 & 5.09 &  7.3e+11 &  1.277e+04 & 12.11 & 7.90e-02 & 5.52e-02 \\ 
5.00 & 10.00 & 2.54 & 1000 & 5.09 &  7.3e+11 &  3.194e+04 & 12.11 & 1.97e-01 & 1.38e-01 \\ 
\enddata
\tablecomments{The left-hand-side part of the Table are the input parameters, where the right-hand-side are the output time delay properties.\\
$\emph{Column 1:}$ Lens redshift.\\
$\emph{Column 2:}$ Source redshift.\\
$\emph{Column 3:}$ Source angular position, in arcseconds.\\
$\emph{Column 4:}$ 1D velocity dispersion.\\
$\emph{Column 5:}$ Einstein radius, in arcseconds.\\
$\emph{Column 6:}$ Mass enclosed within the critical curves, given the input velocity dispersion.\\
$\emph{Column 7:}$ Time delay, in days. \\
$\emph{Column 8:}$ Change in the time delay over a year's time from transverse motion (see text for more details).\\
$\emph{Column 9:}$ Change in the time delay over a year's time from redshift drift (see text for more details).\\
$\emph{Column 10:}$ Change in the time delay over a year's time from mass increase (see text for more details).\\
}
\end{deluxetable*} 
 
\end{document}